\documentclass[twocolumn,pre,showpacs,amsmath,amssymb,nobibnotes,nofootinbib]{revtex4-1}
\usepackage{epsfig}
\usepackage{amsmath,amssymb,color}
\usepackage[cp1251]{inputenc}
\usepackage[english]{babel}

\parskip=\medskipamount

\newcommand{\eq}[1]{\eqref{#1}}
\newcommand{\fig}[1]{Fig.~\ref{#1}}

\newcommand{\be}{\begin{equation}}
\newcommand{\ee}{\end{equation}}

\newcommand{\barr}{\begin{array}}
\newcommand{\earr}{\end{array}}

\newcommand{\ben}{\begin{eqnarray}}
\newcommand{\een}{\end{eqnarray}}

\newcommand{\bs}{\begin{subequations}}
\newcommand{\es}{\end{subequations}}

\newcommand{\bw}{\begin{widetext}}
\newcommand{\ew}{\end{widetext}}

\newcommand\disp{\displaystyle}

\newcommand{\la}{\left<}
\newcommand{\ra}{\right>}

\begin{document}

\title{Phase transition in random planar diagrams and RNA-type matching}

\author{Andrey Y. Lokhov$^{1}$, Olga V. Valba$^{1,2}$, Mikhail V. Tamm$^{3}$, and Sergei K. Nechaev$^{1,4}$}
\affiliation{$^{1}$Universit\'e Paris-Sud/CNRS, LPTMS,
UMR8626, B\^at. 100, 91405 Orsay, France,}
\affiliation{$^{2}$Moscow Institute of Physics and Technology, 141700, Dolgoprudny,
Russia,}
\affiliation{$^{3}$Physics Department, Moscow State University, 119992, Moscow, Russia,}
\affiliation{$^{4}$P.N. Lebedev
Physical Institute of the Russian Academy of Sciences, 119991, Moscow, Russia.}
\date{\today}


\begin{abstract}
We study the planar matching problem, defined by a symmetric random matrix with independent
identically distributed entries, taking values 0 and 1. We show that the existence of a perfect
planar matching structure is possible only above a certain critical density, $p_{c}$, of allowed
contacts (i.e. of '1's). Using a formulation of the problem in terms of Dyck paths and a matrix
model of planar contact structures, we provide an analytical estimation for the value of the
transition point, $p_{c}$, in the thermodynamic limit. This estimation is close to the critical
value, $p_{c} \approx 0.379$, obtained in numerical simulations based on an exact dynamical
programming algorithm. We characterize the corresponding critical behavior of the model and discuss
the relation of the perfect-imperfect matching transition to the known molten-glass transition in
the context of random RNA secondary structure's formation. In particular, we provide strong
evidence supporting the conjecture that the molten-glass transition at $T=0$ occurs at $p_{c}$.
\end{abstract}
\maketitle

\section{Introduction}

\begin{figure}[ht]
\centering
\epsfig{file=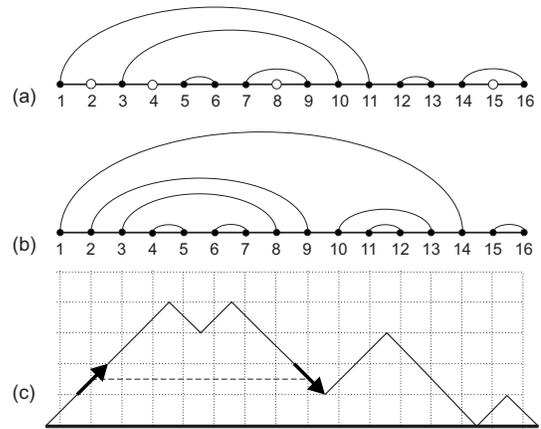,width=7cm} \medskip \caption{Examples of imperfect (a) and
perfect (b) planar matchings (open dots represent gaps); the gapless representation is shown by a
Dyck path (c). The arc is given by an ``up'' and ``down'' steps at the same height, shown by arrows
$\nearrow$ and $\searrow$. The part of the walk between arrows is a Dyck path itself.}
\label{fig:02}
\end{figure}

In this paper the combinatorial problem of complete planar matching is considered. It can
be formulated as follows. Take $L$ points $i=1, \ldots, L$ on a line, and define a symmetric
$L\times L$ random matrix $A$ containing '1's or '0's. We are looking for a set of $L/2$ links
between pairs of points allowed by the entries $A_{ij}$ (each point is involved in one link only)
such that these links form a planar diagram (cf.~\fig{fig:02}). This problem can be thought of as a
constraint satisfaction problem (CSP) characterized by a certain distribution $P(A_{ij})$ on the
entries of the matrix $A$. If at least one such set of links exists, we say that the problem is
satisfiable, and we refer to the solution as to the \emph{perfect matching} configuration.

The Bernoulli model that we study in this paper is defined as follows: $A_{ij}=A_{ji}$ are
independent identically distributed random variables, equal to one with probability $p$ for any
$i\neq j$, and equal to zero otherwise. This sets the uniform distribution on the entries of the
matrix $A$ ($A_{ij}=A_{ji}$):
\be
{\rm Prob}(A_{ij})=p \delta(A_{ij}-1) + (1-p) \delta(A_{ij}),
\label{eq:1-1}
\ee
where $\delta(x)=1$ for $x=0$, and $\delta(x)=0$ otherwise.

The matrix $A$ can be regarded as an adjacency matrix of a random Erd\"os-R\'enyi graph $G(A)$
without self-connections. In what follows, we describe a phase transition, typical for CSP's. A
well-known example of such transition is the SAT-UNSAT problem \cite{Friedgut1999}. As the number
of constraints per node, imposed by the matrix $A$, is below some certain critical value, $p_{c}$,
the problem is satisfiable, while above $p_{c}$ it becomes unsatisfiable in the large $L$ limit. In
other words, there is a critical value of the bond formation probability, $p_{c}$, such that for
any large ($L\gg 1$) sample of the matrix $A$, for $p>p_{c}$ it is always possible to find at least
one ``gapless'' planar diagram, which involves in its formation almost all vertices and only $\sim
\!o(L)$ vertices are missing, while for $p<p_{cr}$ a finite fraction of missing vertices of order
$\sim\!O(L)$ exists, see~\fig{fig:02}(a),(b).

To the best of our knowledge, in the context of random matrix theory, this transition has never
been discussed in the literature. The planar diagrams play an important role in various branches of
theoretical physics, such as matrix and gauge theories \cite{Brezin1978}, many-body condensed
matter physics \cite{AbrikosovGorkov1975}, quantum spin chains \cite{Saito1990} and random matrix
theory \cite{Mehta2004}. Since we do not assume the sparsity of the matrix $A$, the spectral
techniques developed for the sparse random matrices \cite{RodgersBray1988,Mirlin1991,
SemerjianCugliandolo2002} seem to be not applicable in the present problem.

Besides of the mathematical context, this problem has a straightforward application to finding the
optimal secondary structure in RNA molecules \cite{DeGennes1968,Erukhimovich1978, Nussinov1980,
BundschuhHwa1999,Pagnani2000,MontMezard2001, BundschuhHwa2002,OrlandZee2002, Mueller2003,
LassigWeise2006,HuiTang2006,TammNechaev2007}. The secondary RNA structure consists of a set of
saturating reversible chemical bonds between monomers. These bonds correspond to links represented
by planar diagrams. In the RNA context, the matrix $A$ is a function of a frozen disorder in the
sequence of monomers (nucleotides). Recently, the existence of a matching transition as a function
of the number of different monomer types (the nucleotide alphabet size) has been demonstrated in
\cite{ValbaTammNechaev2012}. The main contribution of our current work is two-fold. On one hand, it
consists in the description of the ``perfect-imperfect'' phase transition and the determination of
the threshold value $p_{c}$, using both analytic methods and numerical computations. On the other
hand, we treat the relation between a zero-temperature perfect-imperfect matching transition in
random planar diagrams, and a temperature-dependent molten-glass transition in random RNAs, widely
discussed in the literature \cite{Pagnani2000,BundschuhHwa2002,Marinari2002,KrzakalaMezard2002}. We
find that the perfect-imperfect phase transition point lies on the critical line, separating molten
and glassy regions, and coincides with the freezing transition at zero temperature. Therefore,
while on the corresponding phase diagram the molten phase exists in both perfect and imperfect
regions, the glassy phase is present only in the region with gaps.

The paper is organized as follows. First, we give an estimation of the transition value $p_{c}$ by
mapping of the problem to the so-called Dyck paths and estimating the fraction of ``essential''
arcs. Then, we formulate the planar matching problem in terms of the matrix field theory proposed
in \cite{VernizziOrlandZee2005} and discuss the self-consistent mean-field approximation. The
estimations of the critical point, $p_c$, obtained analytically are compared to the values computed
numerically via an exact dynamical programming algorithm. Finally, we characterize the fluctuation
behavior of the free energy and discuss the relation between the perfect-imperfect transition and
the molten-glass phase transition in random RNA with quenched primary sequence.

\section{Mapping on Brownian excursions and naive mean-field}

An intuitive idea about the calculation of the critical value $p_{c}$ can be obtained by
considering the one-to-one mapping between the $L$-point planar diagrams and the $L$-step Brownian
excursions (BE), known as Motzkin paths \cite{Lando2003} (these paths are also called ``height
diagrams'' in the context of applications to RNA folding \cite{BundschuhHwa2002}). Within this
mapping, the gapless (perfect) planar configurations correspond to BE with no horizontal steps,
also known as Dyck paths, cf. \fig{fig:02}(b),(c). The total number of Dyck paths of even length $L$ is
given by a Catalan number
\be
C_{L/2}=\frac{L!}{(\frac{L}{2})!(\frac{L}{2}-1)!} \sim L^{-3/2}\, 2^L,
\label{catalan}
\ee
where the asymptotic expression is valid for $L \gg 1$; $C_{L/2}$ represents the number of possible
planar diagrams in the \emph{fully-connected} case, corresponding to $p=1$ in our model.

For $0<p<1$, some planar diagrams in the fully-connected ensemble are forbidden. This reduces the
number of possible planar configurations, which becomes zero below a certain value $p_{c}$. A naive
estimation of $p_c$ can be easily obtained via the following mean-field-like argument. Since each
arc in the diagram is present with a probability $p$, the probability that the whole configuration
is allowed, is given by $p^{L/2}$. Assuming that planar diagrams in the fully-connected ensemble
are \emph{statistically independent}, we get the probability $\mathcal{P}$ to have at least one
perfect planar matching configuration:
\be
\mathcal{P}=1-(1-p^{L/2})^{C_{L/2}}=1-\exp\left(-p^{L/2}C_{L/2}\right),
\ee
where the last equality is valid for $L \to \infty$. In this limit, the probability $\mathcal{P}$
is equal to one for $p>p_{c}$, and to zero for $p<p_{c}$. The perfect-imperfect naive  mean-field
threshold $p_{c}$ is thus given by the condition
\be
\lim_{L \to \infty} p_c \left[C_{L/2}\right]^{2/L} = 1,
\label{eq:2-1}
\ee
yielding $p_c=1/4$. However, here we have neglected the statistical correlations between different
configurations in the fully-connected ensemble of planar configurations. Therefore, it provides
only a lower bound to the true value of $p_c$. A careful account for such correlations leads to a
natural generalization of the critical condition \eq{eq:2-1}:
\be
\lim_{L \to \infty} \xi(p_c) \left[C_{L/2}\right]^{2/L} = 1, \; \xi(p_c) = 1/4,
\label{eq:2-2}
\ee
where $\xi(p)$ is some weight (due to correlations) to be determined.

\section{Combinatorics of ``corner counting''}

An estimation of $\xi(p)$, and therefore of $p_c$, can be obtained by exploiting the combinatorial
properties of Dyck paths. The consideration below provides an intuitive understanding of the
statistical reasons beyond the shift of the transition probability from the mean-field value
$p_c=1/4$.

Our estimation is based on the following observation: the probabilities to find different arcs in a
perfect matching structure crucially depend on the lengths of arcs. Consider a perfect matching of
$L/2$ arcs connecting $L$ points. In the limit $L \to \infty$ the \emph{local statistics} of ``up''
and ``down'' steps in a corresponding Dyck path becomes independent on the global constraint for
the random walk to be a Brownian excursion. Using the bijection between Dyck paths and arc
diagrams, we see that the arc is drawn between $i$th and $j$th steps if and only if the $i$th step
is $\nearrow$ (``up'') and $j$th step is the first step $\searrow$ (``down'') at the same height
after $i$ -- as it is depicted in the \fig{fig:02}(c). Therefore, the probability to find an arc from
$i$ to $j$ in a randomly chosen diagram can be formally written as a ``correlation function'':
\be
P(i,j)=\frac{\la\nearrow | \mathcal{D}_{i+1,j-1}| \searrow\ra}{2^{j-i+1}}.
\label{eq:2-3}
\ee
In this expression, the denominator represents the total number of possible sequences from $i$th to
$j$th step; $\mathcal{D}_{i+1,j-1}$ is a Dyck path between $(i+1)$th to $(j-1)$th steps: this part
of the walk should be a Dyck path itself to return to the same spatial coordinate for the
\emph{first time} at $j$th step. The number of such Dyck paths is given by the Catalan number
$C_{(j-i-1)/2}$. Thus, $P(i,j)$ depends only on $k=j-i$ and equals to
\be
P(i,j)=\frac{C_{(k-1)/2}}{2^{k+1}},
\label{eq:2-3a}
\ee
they are non-zero for odd $k$ only: $P(i,i+1)=1/4$, $P(i,i+3)=1/16$, $P(i,i+5)=1/32$,~\emph{etc}.
The whole set of $P(i,j)$ sums to $\sum_{k=1}^{\infty} P(i,i+k)=1/2$, which has a meaning of a
probability that $i$ is a starting (rather than ending) point of an arc.

Thus, the fraction of the shortest arcs of length $k=1$, $P(i,i+1)=1/4$, represented by ``up
corners'' $\wedge$ in a Dyck path, is exceptionally high. Indeed, in a typical fully connected
diagram one half of the arcs ($L/4$ out of $L/2$) correspond to such corners. Moreover, while a
fraction of long arcs chosen in each particular diagram is decaying at $L\to\infty$ (the number of
possible long arcs is of order $L^2$, so the fraction of those chosen in each structure, is of
order $L^{-1}$), the fraction of corners converges to $1/4$ (there are $L-1$ possible corners,
$L/4$ of them appear in a typical structure). Therefore, the values of quenched weights $A_{i,i+1}$
assigned to the short arcs in our model influence the existence of a perfect arc structure in a
crucial way. Now we estimate how this exceptional role of the sub-diagonal values $A_{i,i+1}$
influences $p_c$.

\begin{figure}[ht]
\centering \epsfig{file=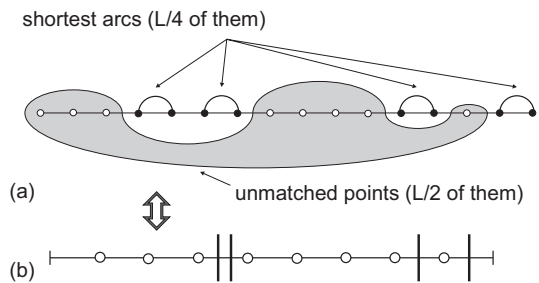,width=7cm}
\caption{Computation of ${\cal Z}$ and ${\cal Z}(p)$:
(a) Selection of $L/4$ non-touching arcs on the set of $L$ points ($L/2$ open dots remain
unmatched); (b) the same problem reformulated as a partitioning of vertical  segments (arcs)
between open dots (unmatched points). A certain number of partitions are forbidden by the matrix of
contacts $A$.}
\label{fig:03}
\end{figure}

Assume that the typical arc structure is constructed as follows: i) take $1/4$ corners (from $L-1$
possible places) such that none of them touch each other, ii) select remaining $L/2-L/4 = L/4$ arcs
at random from ensemble of \emph{any longer} arcs. Since the total number of longer arcs is of
order of $L^2 \gg L/4$, we assume that the quenched disorder in the entries $A_{i,j}$ away from the
sub-diagonal can be safely ignored, and the contribution from the longer arcs into $\xi(p)$ remains
as it is in the mean-field case (each arc is allowed with a probability $p$ independently of
others), thus
\be
\xi^{L/2}(p)= \underbrace{p^{L/4}}_{\rm long\;arcs}\, \underbrace{{\cal P}_{\wedge}(p)}_{\rm
corners}\,,
\label{eq:2-4}
\ee
To determine the contribution of corners, ${\cal P}_{\wedge} (p)$, note that this value has a
meaning of a probability to take $L/4$ corners at random (respecting the non-touching constraint)
in a way that all of them belong to the set of $pL$ allowed ones. Due to the non-touching
constraint the problem of distributing corners can be mapped onto a problem of choosing $L/4$
objects (corners) out of $3L/4$ ones ($L/4$ corners and $L- 2 \times L/4 = L/2$ unmatched vertices,
see \fig{fig:03}). The number of corresponding partitions ${\cal Z}$ is
\be
{\cal Z}=C^{L/4}_{3L/4}=\frac{\frac{3L}{4}!}{\frac{L}{4}!\frac{L}{2}!}
\label{eq:z}
\ee

In the Bernoulli model, only the fraction $p$ of arc positions is fixed. Because of the
non-touching constraint, it is natural to assume that of $3L/4$ positions in the
``point-and-stick'' representation in \fig{fig:03}(b) only $p(L-L/4)=3pL/4$ are allowed on average
(i.e., correspond to unity weights in the connectivity matrix $A$). Thus, the number of allowed
partitions can be estimated as
\be
{\cal Z}(p) =C^{L/4}_{3pL/4}=\frac{\frac{3pL}{4}!}{\frac{L}{4}!(\frac{3pL}{4}-\frac{L}{4})!}.
\label{eq:z(p)}
\ee
Here ${\cal Z}(p)$ is the average number of possibilities to distribute shortest non-touching arcs
at a given fraction $p$ of allowed arcs, and
\be
{\cal P}_{\wedge}(p)=\frac{{\cal Z}(p)}{{\cal Z}}
\label{eq:P_corn}
\ee
is a probability, given $p$, to pick an allowed set of short arcs at random. Substituting
\eq{eq:P_corn} into \eq{eq:2-4} one gets in the limit $L\to\infty$ the following result for
$\xi(p)$:
\be
\ln \xi(p) = \frac{1}{2} \ln p + \frac{3p}{2} \ln\frac{3p}{2} - \frac{3p-1}{2} \ln \frac{3p-1}{2} -
\frac{3}{2} \ln \frac{3}{2}.
\label{eq:2-5}
\ee
Being substituted into \eq{eq:2-2}, Eq. \eq{eq:2-5} gives an estimate for the transition point
$p_{c} = 0.35$. We see therefore that the transition point shifts significantly from its
mean-field value due to the special role of a quenched disorder in the sub-diagonal entries
$A_{i,i+1}$ of the connectivity matrix $A$.

\section{Self-consistent field theory for planar arc counting}

A different way to attack the planar matching problem consists in using the matrix model approach
proposed in \cite{VernizziOrlandZee2005} and the $1/N$-expansion, a standard technique in quantum
field theory. For the set of $L$ vertices, associate to a vertex $i$ an Hermitian matrix
$(\phi_i)_{N \times N}$. The $L$-point generating function $Z_{L}$ can be written as follows:
\be
Z_{L}(N;A) = \frac{\disp \int d\phi_1...d\phi_L e^{-H_{0}} \frac{1}{N} {\rm tr}
\left( \phi_1...\phi_L \right)}{\disp \int d\phi_1...d\phi_L e^{-H_{0}}} \equiv \disp \la
\phi_1...\phi_L \ra_{H_{0}}
\label{eq:3-1a}
\ee
where
\be
H_{0} = \frac{N}{2} \sum_{i,j} (A^{-1})_{ij} {\rm tr}(\phi_i\phi_j).
\label{eq:3-2a}
\ee
Since ${\rm tr} ( \phi_{i} \phi_{j} ) = \sum_{a,b}\phi^{i}_{ab}\phi^{j}_{ba} = \sum_{a,b,c,d}
\delta_{ad}\delta_{bc} \phi^{i}_{ab}\phi^{j}_{cd}$, every propagator enters with a
$1/N$-factor, while every loop gives a $N$--factor. Due to the Wick theorem, one has:
\be
\la \phi_1 \ldots \phi_L \ra_{H_{0}} = \la \sum_{{\rm
pairs}} \prod_{k,k'} \phi_k \phi_{k'} \ra_{H_{0}}
\label{eq:3-3}
\ee
where each non-planar configuration comes with a factor $1/N^{2}$ to some power and therefore
vanishes in the $N \to \infty$ limit. Thus, the generating function $Z_{L}(N;A)$ counts in the
limit $N\to\infty$ the number of planar diagrams with exactly $L/2$ arcs (on genus $g=0$ surface)
compatible with a specific realization of the disorder defined by the matrix $A$. In the absence of
any disorder, one can set $A_{ij} \equiv \alpha$ for any $(i,j)$, where $\alpha$ is some constant
(it corresponds to the $p=1$ limiting case). In this case the multi-dimensional integral
\eq{eq:3-1a} can be reduced by a series of Hubbard-Stratonovich transformations to a
one-dimensional integral involving the spectral density of a Gaussian matrix, which is a well-known
result of the Random Matrix Theory (RMT). We will refer to this realization of $A$ as to the
fully-connected case. If we set $\alpha=1$, we get \cite{VernizziOrlandZee2005}
\be
\lim_{N\to\infty}Z_{L}(N;A) = C_{L/2},
\label{eq:3-4}
\ee
where $C_{L/2}$ is the Catalan number \eqref{catalan}, as it should be. However, for a generic
disordered matrix $A$, the calculations are intractable. Still, we show below that by averaging
over the matrix distribution \eq{eq:1-1} and by applying the self-consistency arguments, we are
able to treat the partially-connected system with $0<p<1$ as an effective fully-connected system
with $\alpha$ different from one, thus obtaining a correction to the naive mean-field result
\eq{eq:2-1}.

According to the consideration above, the function $\xi(p)$ defined by Eq. \eq{eq:2-2} can be
calculated within the matrix approach by averaging $Z_{L}(N;A)$ over the distribution \eq{eq:1-1}.
To this end, we use the standard Hubbard-Stratonovich transform and integrate over $A$ with the
weight \eq{eq:1-1}:
\be
\begin{array}{l}
\disp \int dA ~P(A)\, Z_{L}(N;A) = \\
\disp C \int \prod_{k=1}^{L} d\phi_{k} \frac{1}{N} \text{tr} \left( \phi_{1}
\ldots \phi_{L} \right) \int \prod_{m=1}^{L} dh_{m} e^{i N \sum_{i}
\text{tr}(h_{i}\phi_{i})} e^{S}
\end{array}
\ee
where $C$ is a constant, $S = S_{0} + V$, and
\begin{align}
S_{0} = & - \frac{pN}{2} \sum_{ij} \text{tr}(h_{i}h_{j}),
\label{eq:3-6a} \\
\notag
V = & ~\frac{p(1-p)N^{2}}{8} \sum_{ij} [\text{tr}(h_{i}h_{j})]^{2} \\
& -\frac{p(1-p)(1-2p)N^{3}}{48} \sum_{ij} [\text{tr}(h_{i}h_{j})]^{3} + \ldots
\label{eq:3-6}
\end{align}

Up to this point, no approximation has been made. The $S_{0}$ term \eq{eq:3-6a} corresponds to a
fully-connected matrix with an additional factor $p$ behind. If this term is the only present,
then, performing the inverse Hubbard-Stratonovich transformation and returning to the functional of
the type \eq{eq:3-1a}, we get $\xi(p)=p$, recovering the value $p_{c}=1/4$ given by the critical
condition \eq{eq:2-1}.

The correction to $p_{c}$ due to the rest of the series $V$ \eq{eq:3-6} can be estimated as
follows. The series given by the action $S$ can be thought of as a Gaussian theory with the
interaction $V$. Since $V$ contains an infinite number of terms, it is impossible to treat it
perturbatively. Still, we can use a self-consistent nonperturbartive approach reminiscent of the
Feynman's variational principle \cite{Feynman1955} in the field theory: as all the fields
$\{h_{i}\}_{i=1, \ldots, L}$ in Eq.\eq{eq:3-6} are equivalent, we assume that the average $\langle
N \text{tr} (h_{i} h_{j}) \rangle_{S_{0}} \equiv U$ is independent on $(i,j)$. Within the adopted
mean-field approximation, the replacement $e^{S} = e^{S_{0}} e^{\langle V \rangle}$ is supposed,
where
\be
\begin{array}{ll}
\langle V \rangle = & \disp \frac{p(1-p)N}{8} U \sum_{ij} \text{tr}(h_{i}h_{j})
\\ & \disp - \frac{p(1-p)(1-2p)N}{48} U^{2} \sum_{ij} \text{tr}(h_{i}h_{j}) + \ldots
\end{array}
\label{eq:3-7}
\ee
Resumming the series \eq{eq:3-7}, we obtain the following self-consistent equation for the
``propagator'' $U$:
\be
\frac{1}{U} = -\frac{2}{U} \log{\left[ 1 - p + p \exp{ \left( - \frac{U}{2} \right) } \right] }.
\label{eq:3-8}
\ee
The equation \eq{eq:3-8} yields $U=-2 \log{ \left[ 1 - \frac{1-1/\sqrt{e}}{p} \right] }$. Hence,
finally, we can write
\be
S = - \frac{\xi(p)N}{2} \sum_{ij} \text{tr}(h_{i}h_{j})
\label{eq:3-9}
\ee
where
\be
\xi(p) = \left( -2 \log{ \left[ 1 - \frac{1-1/\sqrt{e}}{p} \right] } \right)^{-1}.
\label{eq:3-10}
\ee
Substituting \eq{eq:3-10} into \eq{eq:2-2}, we get an estimation for the critical value
$p^{*}_{c}=0.455$. Although the self-consistent approximation \eq{eq:3-7} seems to be rather crude
(the numerical estimation of $p_c$ for large matrices is $p_c\approx 0.379$), it leads to the
correct direction of the shift of $p_c$ from the naive mean--field value $p_c=0.25$ \eq{eq:2-1}. It
would be interesting to understand how to treat the interaction term $V$ \eq{eq:3-6} more properly.

\section{Perfect-imperfect phase transition}

The combinatorial problem of planar matching, introduced above, is strongly related to the problem
of optimal folding of random RNAs. A real RNA represents a single-stranded polymer, composed of
four types of nucleotides: A, C, G and U. Under normal conditions, the RNA molecule folds onto
itself and forms a double-helical structures of stacked base pairs, known as the \emph{secondary
structure} of the RNA, favoring the stable Watson-Crick pairs A-U and G-C. The simplest theories
describe the statistics of the \emph{random} RNA secondary structures, incorporating the most
important features: saturation of base-pairings, exclusion of the so-called pseudoknots, that are
known to be very rare in real RNA \cite{Nussinov1980}, and, sometimes, the condition of finite
flexibility of the molecule, requiring a minimal length of a loop \cite{Marinari2002,
KrzakalaMezard2002}. The exclusion of the pseudoknots means that the base pairings can be
represented by one-dimensional planar diagrams, depicted in \fig{fig:02}. This topological
constraint allows to calculate the partition function of the RNA, using an exact dynamical
programming algorithm \cite{Nussinov1980,DeGennes1968}. The recursion relation for the partition
function, $Z_{i,i+k}$, of the part between monomers $i$ and $i+k$, reads:
\be
Z_{i,i+k}=Z_{i+1,i+k}+\sum_{s=i+1}^{i+k} \beta_{i,s}Z_{i+1,s-1}\, Z_{s+1,i+k}
\label{eq:3-1}
\ee
where $\beta_{i,j}=e^{-A_{i,j}/T}$ are statistical weights of bonds ($1\le i < j \le n$);
$A_{i,j}=1$ if $i$ and $j$ match each other, and $A_{i,j}=0$ otherwise. In the zero-temperature
limit, the equation \eq{eq:3-1} is reduced to the dynamical programming algorithm for the ground
state free energy \cite{ValbaTammNechaev2011}:
\be
\begin{array}{l}
\disp F_{i,i+k} = \lim\limits_{T \to 0} T\ln Z_{i,i+k} = \max_{s=i+1,...,i+k} \Big\{F_{i+1,i+k}, \\
\disp \hspace{1.2cm} \varepsilon_{i,s}+F_{i+1,s-1}+F_{s+1,i+k}\Big\}
\end{array}
\label{eq:3-2}
\ee
Since the free energy, $F$, of the whole chain is proportional to the number of nucleotides
involved in the planar bond formation, the combinatorial problem of planar matching can be regarded
as a $T=0$ optimization problem for the free energy of the RNA molecule with a given matrix of
contacts, $A$. Therefore, the exact dynamical programming algorithm \eq{eq:3-2} allows to detect
the phase transition by considering the fraction $f_{L}(p)=2F/L$ of links, involved in planar
binding, for different densities of contacts $p$ in the limit $L \rightarrow \infty$: one expects
$f_{\infty}(p)=1$ for $p>p_{c}$, and $f_{\infty}(p)<1$ for $p<p_{c}$.

\begin{figure}[ht]
\centering
\epsfig{file=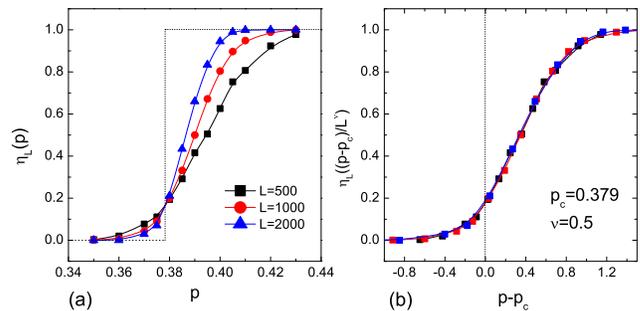,width=8.4cm}
\caption{(Color online) (a) The fraction of perfect matchings $\eta_{L}(p)$ as a function of the density $p$ of
ones in the contact matrix $A$ for chain lengths $L=500,\,1000,\,2000$, averaged over 10000
instances. The dashed line corresponds to the thermodynamic limit $L\to\infty$, yielding the
critical value $p_{c}=0.379$. (b) The scaling analysis of curves, corresponding to different chain
lengths $L$. The fitting procedure gives the exponent of the transition width $\nu \approx 0.5$.}
\label{fig:04}
\end{figure}

Thus, looking for the fraction, $\eta_{L}(p)$, of sequences, which allow perfect matchings, in the
whole ensemble of random sequences, one has $\eta_{\infty}(p)=1$ for $p>p_{c}$, and
$\eta_{\infty}(p)=0$ for $p<p_{c}$. The corresponding dependencies are shown in \fig{fig:04}(a) for
different polymer lengths, $L=500,\,1000,\,2000$. As $L\to \infty$, the function $\eta_{L}(p)$
tends to a step function. Two different phases are observed: for $p>p_{c}$ one has a gapless
perfect matching with all nucleotides involved in planar binding, while for $p<p_{c}$ there is
always a finite fraction of gaps in the best possible matching.

\begin{figure}[ht]
\epsfig{file=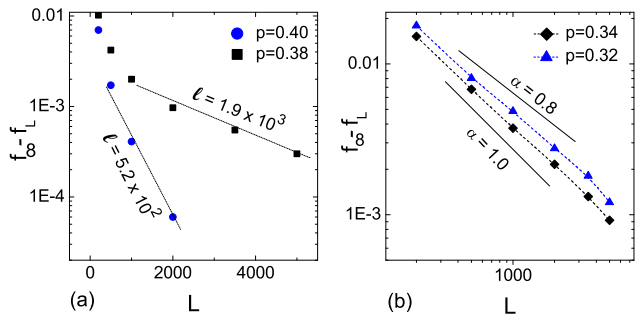,width=8.4cm}
\caption{(Color online) Convergence of fraction of links, involved in planar binding, $f_{L}$ to the
limiting value $f_{\infty}$ in two regimes, $p>p_c$ and $p<p_c$. (a) In the perfect phase, the
exponential convergence is demonstrated for $p=0.38$ and $p=0.4$ in the semi-logarithmic scale. The
screening length $\ell(p)$ diverges as $p$ approaches the critical value $p_{c}$. (b) In the
imperfect phase, the power-law behavior is shown for $p=0.32$ and $p=0.34$ in the log-log scale.
The exponent $\alpha(p)$ as a function of $p$ takes values between $0.8$ and $1$. The data points
are averaged over 1000 instances.}
\label{fig:04a}
\end{figure}

The scaling analysis permits to determine the phase transition point as $p_c \approx 0.379$
(compare to the predictions of the ``corner counting'' and of the self-consistent field theory).
The \fig{fig:04}(b) shows that curves with different $L$ collapse, demonstrating the scaling
behavior $\eta\left((p-p_c)/L^{\nu}\right)$ with the transition width $L^{-\nu}$, where $\nu=0.5$.
The convergence of the function $f_{L}$ to the limiting value $f_{\infty}(p)$ (cf. \fig{fig:04a})
in the perfect and imperfect phases has, respectively, exponential and power-law tails:
\be
\left\{\begin{array}{ll}
f_{\infty}(p) - f_L(p) \sim e^{-L/\ell(p)} & \mbox{for $p>p_c$} \medskip \\
f_{\infty}(p) - f_L(p) \sim L^{-\alpha(p)} & \mbox{for $p<p_c$}
\end{array} \right.
\label{eq:4-1}
\ee
In the perfect matching phase the screening length $\ell(p)$ diverges at the point $p=p_c$ (two
examples for $p=0.38$ and $p=0.4$ are shown in the \fig{fig:04a}(a) in the semi-logarithmic scale),
while for the imperfect matching the finite-size scaling analysis demonstrates the power-law
behavior with the exponents $0.8 \le \alpha(p) \le 1$ (see \fig{fig:04a}(b) for two examples,
$p=0.32$ and $p=0.34$ in the log-log plot). Note that the exponential scaling for $p>p_c$ may not
be universal (being model-dependent) and is likely to be a feature of the Bernoulli model
\eqref{eq:1-1}, while the power-law behavior for $p<p_c$ appears in other models, e.g. for
integer-valued ``alphabet'' \cite{ValbaTammNechaev2012}.

\section{Molten-glass phase transition}

The investigation of thermodynamic properties of RNA secondary structures has been addressed in a
number of papers \cite{BundschuhHwa1999,Pagnani2000,BundschuhHwa2002,KrzakalaMezard2002,
Marinari2002,LassigWeise2006,HuiTang2006}. Many of them provided numerical and analytical evidence
for existence of a low-temperature glassy phase. In \cite{BundschuhHwa2002} it was shown that in
the high-temperature phase the system remains in the molten phase, characterized by a
homopolymer-like behavior. In the molten phase the disorder is irrelevant, and the binding matrix
elements $A_{ij}$ can be replaced by some effective value $\alpha$. Carrying out the two-replica
calculation, the authors were able to prove that the system exhibits a phase transition from a
high-temperature regime, in which the replicas are independent, to a low-temperature phase, in
which the disorder is relevant and replicas are strongly coupled. The authors numerically
characterized the transition to a glassy phase by imposing a pinch between two bases and measuring
the corresponding energy cost.

Several other works \cite{KrzakalaMezard2002,Marinari2002} used an alternative so-called
$\varepsilon$-coupling method, to investigate the nature and the scaling laws of the glassy phase,
observing the effect of typical excitations imposed by a bulk perturbation. The authors argued that
for the models with non-degenerate ground states, the low-temperature phase is not marginal, but is
governed by a scaling exponent, close to $\theta=1/3$. The explicit numerical studies of the
specific heat demonstrate that molten-glass transition is only a fourth order phase transition
\cite{Pagnani2000}.

Regardless of particular details of models considered in all these works, it is clear that the
existence of the glassy phase is possible only in a sufficiently disordered and frustrated system.
Besides the planarity constraint, shared by all simple models of random RNA, the Bernoulli model is
described by a unique disorder parameter, $p$, that controls the density of allowed contacts. In
this model, the appearance of the glassy phase is impossible above a certain threshold, $p^{*}$.
Indeed, it is well-known that for $p=1/2$ (corresponding to an effective alphabet $c=1/p=2$), there
is no transition to the glassy phase at all, and the system remains always in the molten phase
\cite{Pagnani2000,BundschuhHwa2002}. Below, we present arguments, supporting the hypothesis that
$p^{*}$ is equal to the critical value of perfect-imperfect matching transition, $p_{c}$ discussed
above.

\begin{figure}[ht]
\centering
\epsfig{file=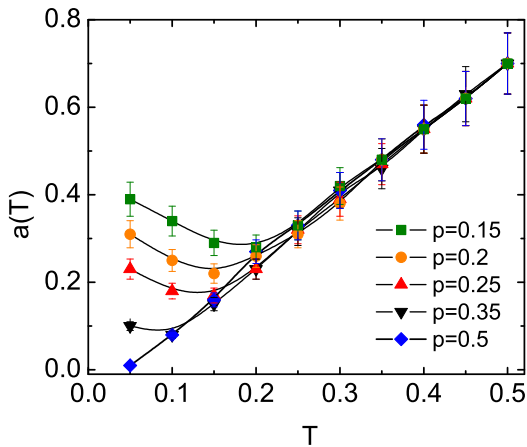,width=7cm}
\caption{(Color online) The dependence of the coefficient $a(T)$ in \eq{eq:4-4} on the temperature for
$p=0.15,0.2,0.25,0.35,0.5$. For $p>p^{*}$ ($0.35<p^{*}<0.5$), the coefficient $a(T)$ seems to
follow the $a(T)=3T/2$ law, typical for the molten phase, up to very low temperatures. For
$p<p^{*}$, the $a(T)$-dependence deviates from the high-temperature behavior at some temperature,
which we identify as a critical temperature of transition to the glassy phase. The data points are
averaged over 10000 samples.}
\label{fig:05}
\end{figure}

To identify the dependence of the molten-glass transition temperature on the effective alphabet
(defined as $c=1/p$), we follow the procedure suggested in \cite{BundschuhHwa2002}. In the
high-temperature regime the disorder is irrelevant (this corresponds to a homopolymer-like behavior
in polymer language) and one can put $A_{ij}=\alpha$. In this regime the free energy of the chain
of length $L$ scales linearly with $L$, up to a logarithmic correction, which is just the logarithm
of the power-law multiplier in the Catalan number \eq{catalan} enumerating all possible structures:
$F(L,T)= f(T) L - (3T/2) \ln L$, where $f(T)$ is some (non-universal) function of the temperature.
In particular, the energy cost of imposing a bond connecting two monomers at distance $L/2$ from
each other equals in the high temperature regime
\be
\Delta F(L,T)= F(L,T) - 2 F (L/2,T) = \frac{3}{2} T \ln \frac{L}{4}.
\label{log_glass}
\ee
The violation of this behavior indicates \cite{BundschuhHwa2002} the appearance of the glassy
phase. This fact can be used to detect the transition temperature in the Bernoulli model. Namely,
we use the following fit for $\Delta F(L,T)$ (where $F(L,T)$ is to be determined via recursion
relations \eq{eq:3-2})
\be
\Delta F(L, T)=a(T)\ln L+b(L),
\label{eq:4-4}
\ee
and plot the $T$-dependence of $a(T)$, see \fig{fig:05}. We interpret the deviation of the $a(T)$
from the high-temperature value $3T/2$ as appearance of the glass transition. Note that the
logarithmic fit \eq{eq:4-4} for the free energy does not give a correct asymptotics at low
temperatures (indeed, the true asymptotics is known to include power-law and logarithm-squared
terms \cite{HuiTang2006}).

As it follows from \fig{fig:05}, the expected behavior \eq{log_glass} is indeed observed at high
temperatures, and is violated at a certain temperature $T_{c}$. Following \cite{BundschuhHwa2002}
we identify this regime change with the molten-glass transition. We see that with the increase of
$p$, the critical temperature $T_{c}$ shifts to lower values, approaching zero for some $0.35 <
p^{*} < 0.5$. At low temperatures, the numerical computations become very time consuming, leading
to the loss of precision in the vicinity of $p^{*}$. However, it seems that the hypothesis
$p^{*}=p_{c}$ still holds: the sequences corresponding to $p>p_{c}$ remain in the molten phase, the
pinching free energy \eq{eq:4-4} has the same dependence even for very low temperatures.

\section{Discussion: matching vs freezing}

\begin{figure}[ht]
\centering
\epsfig{file=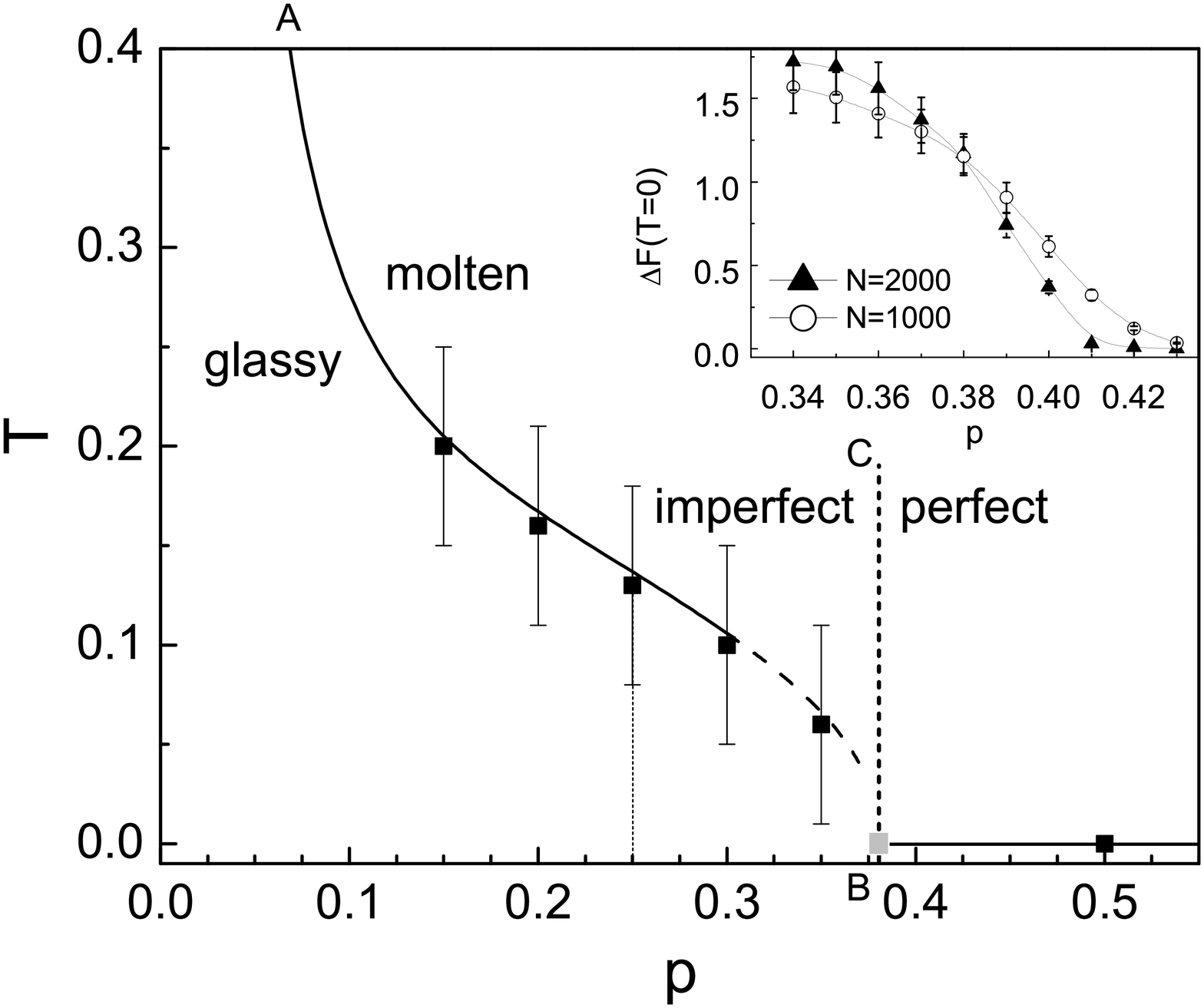,width=7cm}
\caption{Main figure: the phase diagram of Bernoulli model on the $(T,p)$ plane. The data points
correspond to the critical temperature $T_{c}$ of the molten-glass transition for different values
$p=0.15,0.2,0.25,0.3,0.35,0.5$. A 4-letter alphabet ($p=0.25$), is highlighted by a thin dashed
line. The critical curve (A-B) separates glassy and molten phases. We conjecture that at zero
temperature, the endpoint B, giving $p^{*}$, coincides with the critical point $p_{c}$ for the
perfect-imperfect transition. The thick dashed line (B-C) separates the perfect and imperfect
matching cases. The glassy phase lies entirely inside the region, characterized by gaps. Inset: an
evidence for the conjecture $p^{*}=p_{c}$. Study of the pinching free energy $\Delta F(L, T)$ at
zero temperature. In the limit of large $L$, the glassy phase is absent for $p>p^{*}$,
characterized by $\Delta F(\infty, 0)=0$. The point $p^{*}$ can be identified as a crossing point
for different $\Delta F(L, 0)$ curves, presented here for $L=1000$ and $L=2000$, and it's value is
found to be very close to $p_{c}=0.379$. The data points are averaged over 1000 samples.}
\label{fig:06}
\end{figure}

The results presented in this work suggests the generic phase diagram shown in the \fig{fig:06} for
the Bernoulli model of random RNA chains. The perfect-imperfect transition at zero temperature,
separates two matching regions: with and without gaps. Analytically, we proved the existence of the
transition from the perfect matching region to the imperfect one, and provided estimates for the
values of the transition point, $p_{c}$. Using the exact dynamical programming algorithm
\eq{eq:3-2}, we found this critical value to be $p_{c}\approx 0.379$, highlighted by a thick dashed
line (B-C) in \fig{fig:06}. The previous studies have been mostly concentrated on the description
of the finite-temperature molten-glass transition for a sufficiently frustrated model with a fixed
alphabet (a fixed $p$ in the Bernoulli model). An example of such a phase transition point is
marked by a thin dashed line in the \fig{fig:06}, and corresponds to an intensively studied case of
the 4-letter alphabet ($p=0.25$). The ensemble of critical points for different values of $p$ gives
a critical curve (A-B) in the $(T,p)$ plane.

The computational cost increases drastically for temperatures close to zero (and, hence, in the
vicinity of $p_c$), and the recursive relations \eq{eq:3-1} are no more applicable. However, we can
still try to carry out the analysis of the pinching free energy $\Delta F(L, T)$ at zero
temperature, using the exact dynamical programming algorithm \eq{eq:3-2}. Indeed, the glassy phase
does not exist if $\Delta F(\infty, 0)=0$. This happens for $p>p^{*}$, where $p^{*}$ is defined as
the density of constrains, for which the critical temperature is zero: $T_c(p^{*})=0$. The
corresponding plot is shown in the inset of \fig{fig:06}. According to \eq{eq:4-1}, the pinching
free energy \eq{log_glass} decreases with growth of $L$ in the imperfect matching phase, while
increases (with growth of $L$) in the perfect matching regime. Hence, the value of $p^{*}$ in the
large $L$ limit can be identified as a crossing point of $\Delta F(L, 0)$ curves for different $L$.
The crossing point for $L=1000$ and $L=2000$ is indeed found to be very close to the value
$p_{c}=0.379$, strongly supporting the hypothesis $p^{*}=p_{c}$. The aforementioned results
indicate that the critical curve $T_c(p)$ crosses zero at the critical value $p_{c}$. Hence, the
perfect-imperfect transition point seems to lie at the critical line, separating molten and glassy
regions, and coincides with its limiting $T=0$ value. We see that although the glassy phase exists
only in the region where the gaps are present, the molten phase lies in both, perfect and
imperfect, matching regions.

Because of the one-parameter dependence, the Bernoulli model is probably the simplest model for
modelling the secondary structure of the RNA, that captures the essential physical properties of
the process. Being applied to the studies of the thermodynamic properties of random RNAs, the
problem introduced in this paper provides some enlightenment on the nature of molten-glass
transition at zero temperature. Starting from Bernoulli model, one could directly generalize our
approach to investigate more sophisticated and realistic models of the RNA secondary structure, for
example, by introducing the minimal allowed hairpin length \cite{Pagnani2000,BundschuhHwa2002,
KrzakalaMezard2002}, taking into accounts the pseudoknots \cite{VernizziOrlandZee2005} and
different binding probabilities \cite{VernizziOrlandZee2005,Marinari2002}.

\begin{acknowledgments}
The authors are grateful to V. Avetisov, M. M\'ezard, V. Stadnichuk and A. Vladimirov for encouraging
discussions and valuable comments. This work was partially supported by the grants
ANR-2011-BS04-013-01 WALKMAT, ERASysBio+ $\#66$ (ANR-09-SYSB-004), FP7-PEOPLE-2010-IRSES 269139
DCP-PhysBio and by a MIT--France Seed fund ``Genome in 3D: Fractal and Topological Properties of
DNA Folding''.

\end{acknowledgments}


\begin{thebibliography}{99}

\bibitem{Friedgut1999}
Friedgut, E. Necessary and sufficient conditions for sharp thresholds and the k-sat problem {\em J.
Amer. Math. Soc.} {\bf 12}, 1017--1054 (1999).

\bibitem{Brezin1978}
Br\'{e}zin, E, Itzykson, C, Parisi, G,  \& Zuber, J.~B. Planar diagrams. {\em Communications in
Mathematical Physics} {\bf 59}, 35--51 (1978).

\bibitem{AbrikosovGorkov1975}
Abrikosov, A.~A \& Gorkov, L.~P. {\em Methods of quantum field theory in statistical physics}.
(Courier Dover Publications) (1975).

\bibitem{Saito1990}
Saito, R. A Proof of the Completeness of the Non Crossed Diagrams in
  Spin 1/2 Heisenberg Model.
{\em Journal of the Physics Society Japan} {\bf 59}, 482--491 (1990).

\bibitem{Mehta2004}
Mehta, M.~L. {\em {Random matrices}}. (Academic press) (2004).

\bibitem{RodgersBray1988}
Rodgers, G \& Bray, A. Density of states of a sparse random matrix. {\em Physical Review B} {\bf
37}, 3557--3562 (1988).

\bibitem{Mirlin1991}
Mirlin, A.~D \& Fyodorov, Y.~V. Universality of level correlation function of sparse random
  matrices.
{\em Journal of Physics A: Mathematical and General} {\bf 24},
  2273--2286 (1991).

\bibitem{SemerjianCugliandolo2002}
Semerjian, G \& Cugliandolo, L.~F. Sparse random matrices: the eigenvalue spectrum revisited. {\em
Journal of Physics A: Mathematical and General} {\bf 35},
  4837--4851 (2002).

\bibitem{DeGennes1968}
de~Gennes, P.~G. Statistics of branching and hairpin helices for the dAT
  copolymer.
{\em Biopolymers} {\bf 6}, 715--29 (1968).

\bibitem{Erukhimovich1978}
Erukhimovich I.Ya. On Size and Some Structural Characteristics of Moderately Cross-Linked Long
Polymer Chains. {\em Vysokomol. Soyed.} {\bf 20B}, 10 (1978).

\bibitem{Nussinov1980}
Nussinov, R \& Jacobsont, A.~B. Fast algorithm for predicting the secondary structure of. {\em
Proceedings of the National Academy of Sciences} {\bf 77},
  6309--6313 (1980).

\bibitem{BundschuhHwa1999}
Bundschuh, R \& Hwa, T. RNA Secondary Structure Formation: A Solvable Model of
  Heteropolymer Folding.
{\em Physical Review Letters} {\bf 83}, 1479--1482 (1999).

\bibitem{Pagnani2000}
Pagnani, A, Parisi, G,  \& Ricci-Tersenghi, F. Glassy Transition in a Disordered Model for the RNA
Secondary Structure. {\em Physical Review Letters} {\bf 84}, 2026--2029 (2000).

\bibitem{MontMezard2001}
Montanari, A. \& Mezard, M. Hairpin Formation and Elongation of Biomolecules {\em Phys. Rev.
Letters} {\bf 86} 2178--2181 (2001).

\bibitem{BundschuhHwa2002}
Bundschuh, R \& Hwa, T. Statistical mechanics of secondary structures formed by
  random RNA sequences.
{\em Physical Review E} {\bf 65}, 031903 (2002).

\bibitem{OrlandZee2002}
Orland, H.  \& Zee, A. RNA folding and large N matrix theory. {\em Nuclear Physics B} {\bf 620},
456--476 (2002).

\bibitem{Mueller2003}
M\"uller M. Statistical physics of RNA folding. {\em Physical Review E} {\bf 67} 021914 (2003).

\bibitem{LassigWeise2006}
L\"{a}ssig, M \& Wiese, K. Freezing of Random RNA. {\em Physical Review Letters} {\bf 96}, 228101
(2006).

\bibitem{HuiTang2006}
Hui, S. \& Tang, L.-H. Ground state and glass transition of the RNA secondary
  structure.
{\em The European Physical Journal B} {\bf 53}, 77--84 (2006).

\bibitem{TammNechaev2007}
Tamm, M.V. \& Nechaev, S.K. Necklace-cloverleaf transition in associating RNA-like diblock
copolymers. {\em Physical Review E} {\bf 75}, 031904 (2007).

\bibitem{ValbaTammNechaev2012}
Valba, O.~V, Tamm, M.~V,  \& Nechaev, S.~K. New Alphabet-Dependent Morphological Transition in
Random RNA  Alignment. {\em Physical Review Letters} {\bf 109}, 018102 (2012).

\bibitem{Marinari2002}
Marinari, E, Pagnani, A,  \& Ricci-Tersenghi, F. Zero-temperature properties of RNA secondary
structures. {\em Physical Review E} {\bf 65}, 041919 (2002).

\bibitem{KrzakalaMezard2002}
Krzakala, F, M\'{e}zard, M,  \& M\"{u}ller, M. Nature of the glassy phase of RNA secondary
structure. {\em Europhysics Letters (EPL)} {\bf 57}, 752--758 (2002).

\bibitem{VernizziOrlandZee2005}
Vernizzi, G, Orland, H,  \& Zee, A. Enumeration of RNA Structures by Matrix Models. {\em Physical
Review Letters} {\bf 94}, 168103 (2005).

\bibitem{Lando2003}
Lando, S.~K. {\em {Lectures on generating functions}}. (Amer Mathematical Society) (2003).

\bibitem{Feynman1955}
Feynman, R. Slow Electrons in a Polar Crystal. {\em Physical Review} {\bf 97}, 660--665 (1955).

\bibitem{ValbaTammNechaev2011}
Nechaev S.~K, Tamm M.~V, \& Valba O.~V. Sequence matching algorithms and paring of noncoding RNAs
{\em J. Phys. A} {\bf 44}, 195001 (2011).

\end{thebibliography}
\end{document}